\documentclass[a4paper]{jpconf1}
\usepackage{graphicx}
\usepackage{bm}
\usepackage{url}
\usepackage{setspace}

\def\pT{p_\mathrm{T}}

\begin{document}
\title{Towards the Little Bang Standard Model}

\author{Ulrich Heinz}

\address{Physics Department, The Ohio State University, Columbus, Ohio 43210, USA}

\ead{heinz@mps.ohio-state.edu}

\begin{abstract}
I review recent progress in developing a complete dynamical model for the evolution of the Little Bang fireballs created in relativistic heavy-ion collisions, and using the model to extract the transport properties and initial density fluctuations of the liquid quark-gluon plasma state of matter of which makes up these Little Bangs during the first half of their lives.
\vspace*{-5mm}
\end{abstract}

\section{The Big Bang and the Little Bangs}
\label{sec1}

\begin{figure}[htb]
\vspace*{-2mm}
\begin{center}
\includegraphics[width=0.75\linewidth]{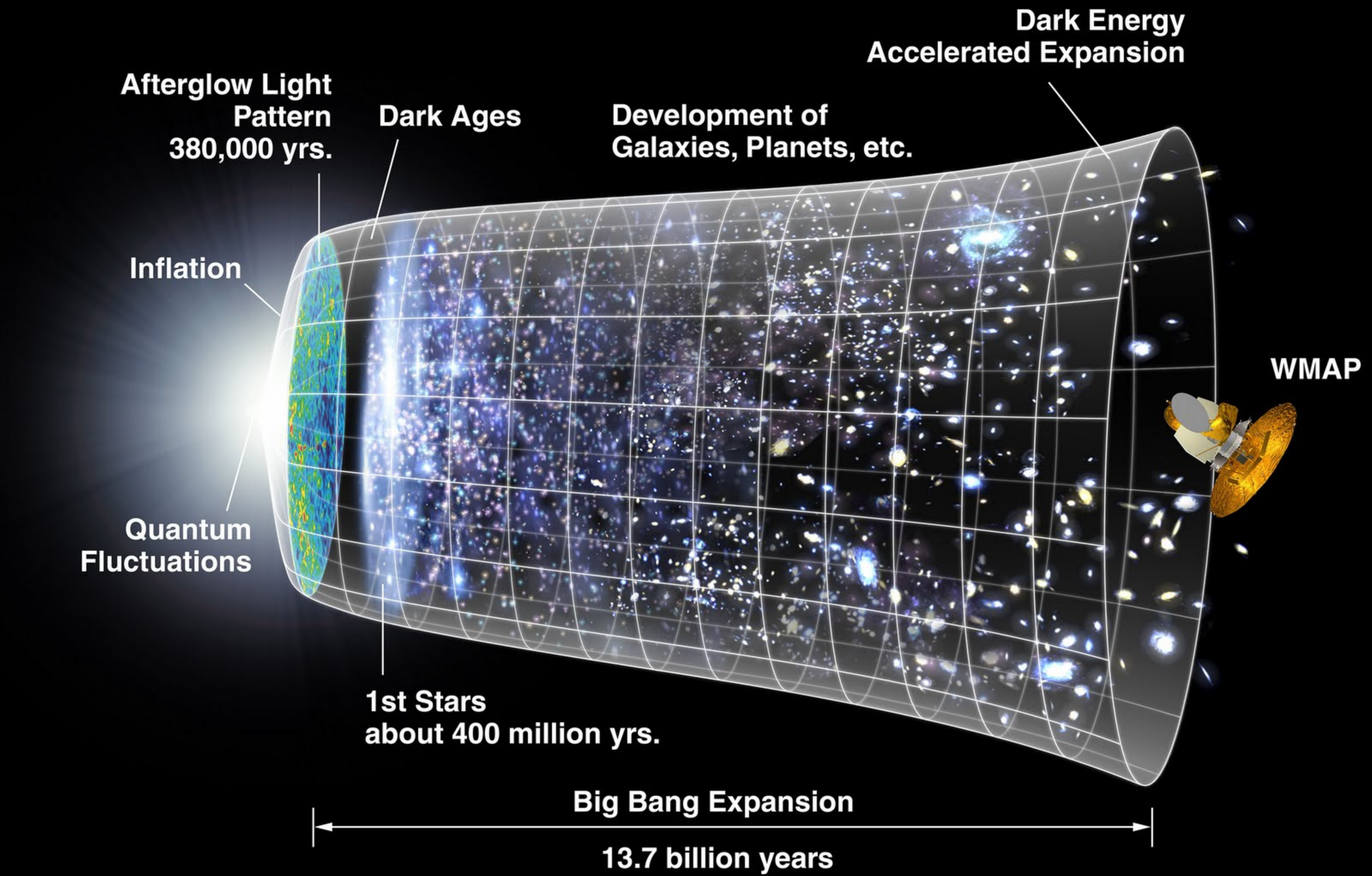}
\includegraphics[width=0.755\linewidth,clip=]{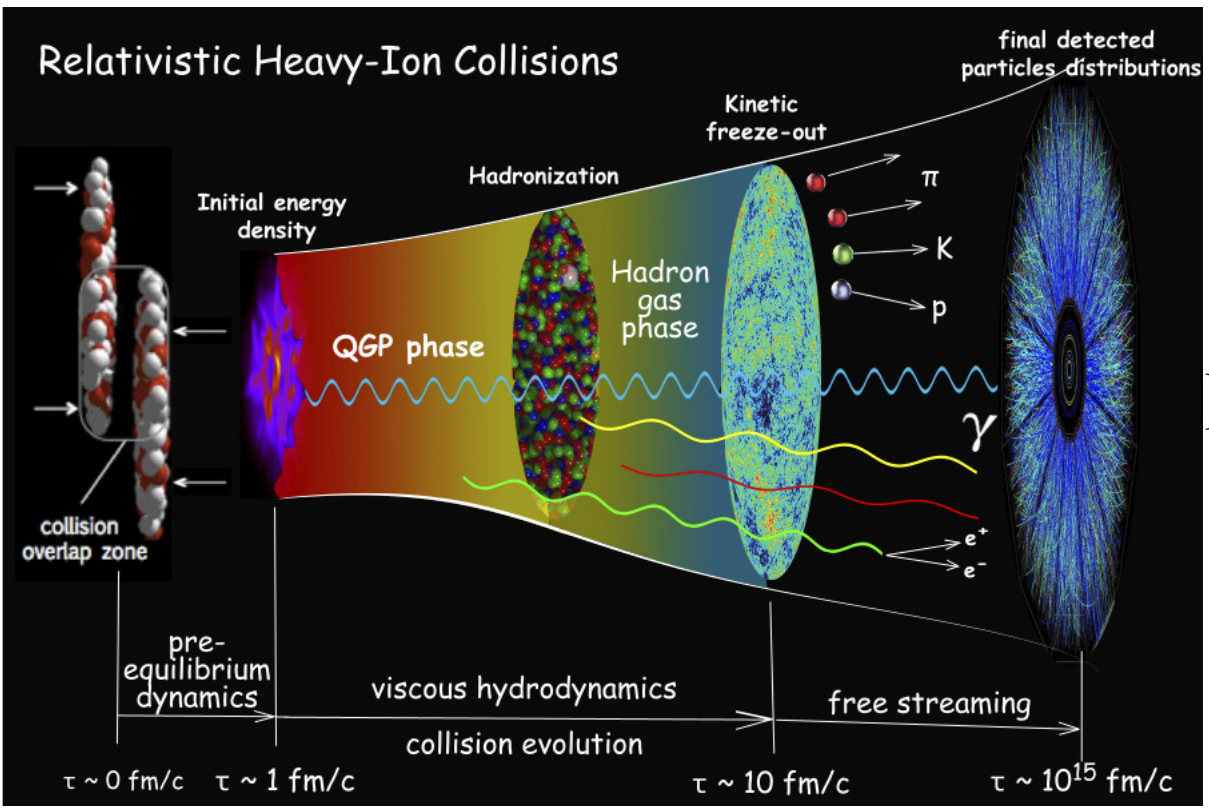}
\end{center}
\begin{spacing}{0.8}
\vspace*{-3mm}
\caption{
\label{F1}
{\footnotesize
Artist's conception of the evolution of the Big Bang (top -- credit: NASA)
and the Little Bang (bottom -- credit: Paul Sorensen and Chun Shen).
}
\vspace*{-5mm}
}
\end{spacing}
\end{figure}

Ultra-relativistic heavy-ion collisions at RHIC and the LHC produce fireballs made of extraordinarily hot matter, at initial energy densities (at the time when the matter reaches approximate local thermal equilibrium) that exceed the energy density of atomic nuclei in their ground states by two to three orders of magnitude. Due to enormous pressure gradients 
%
\begin{figure}[htb]
\vspace*{-2mm}
\begin{center}
\includegraphics[width=0.9\linewidth]{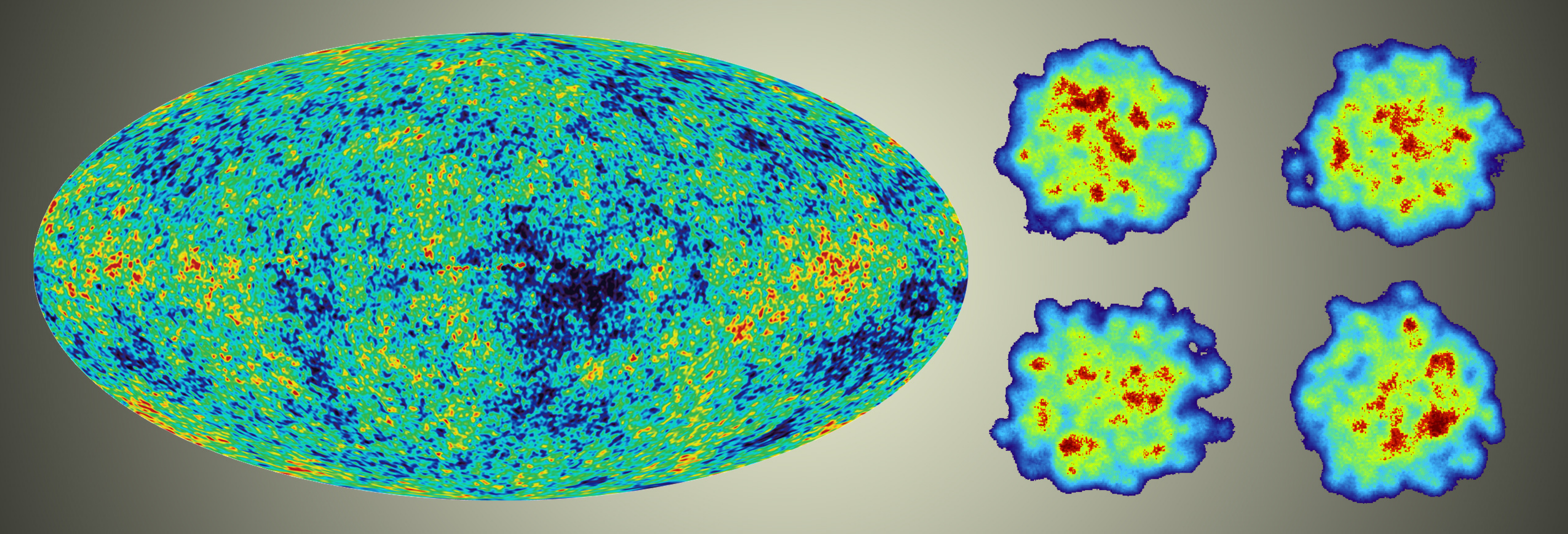}
\end{center}
\begin{spacing}{0.8}
\vspace*{-3mm}
\caption{
\label{F2}
{\footnotesize
Temperature fluctuation spectrum of the Big Bang at age 380,000 yr (left), as measured through the Cosmic Microwave Background radiation by WMAP, and of four typical Little Bangs created in central Pb+Pb collisions at the LHC (right), at age 0.2\,fm/$c$, as calculated from the IP-Glasma model \cite{Schenke:2012wb}. Figure taken from \cite{NSAC}. 
}
\vspace*{-5mm}
}
\end{spacing}
\end{figure}
%
%
\begin{figure}[htb]
\vspace*{-2mm}
\begin{center}
\includegraphics[width=0.85\linewidth]{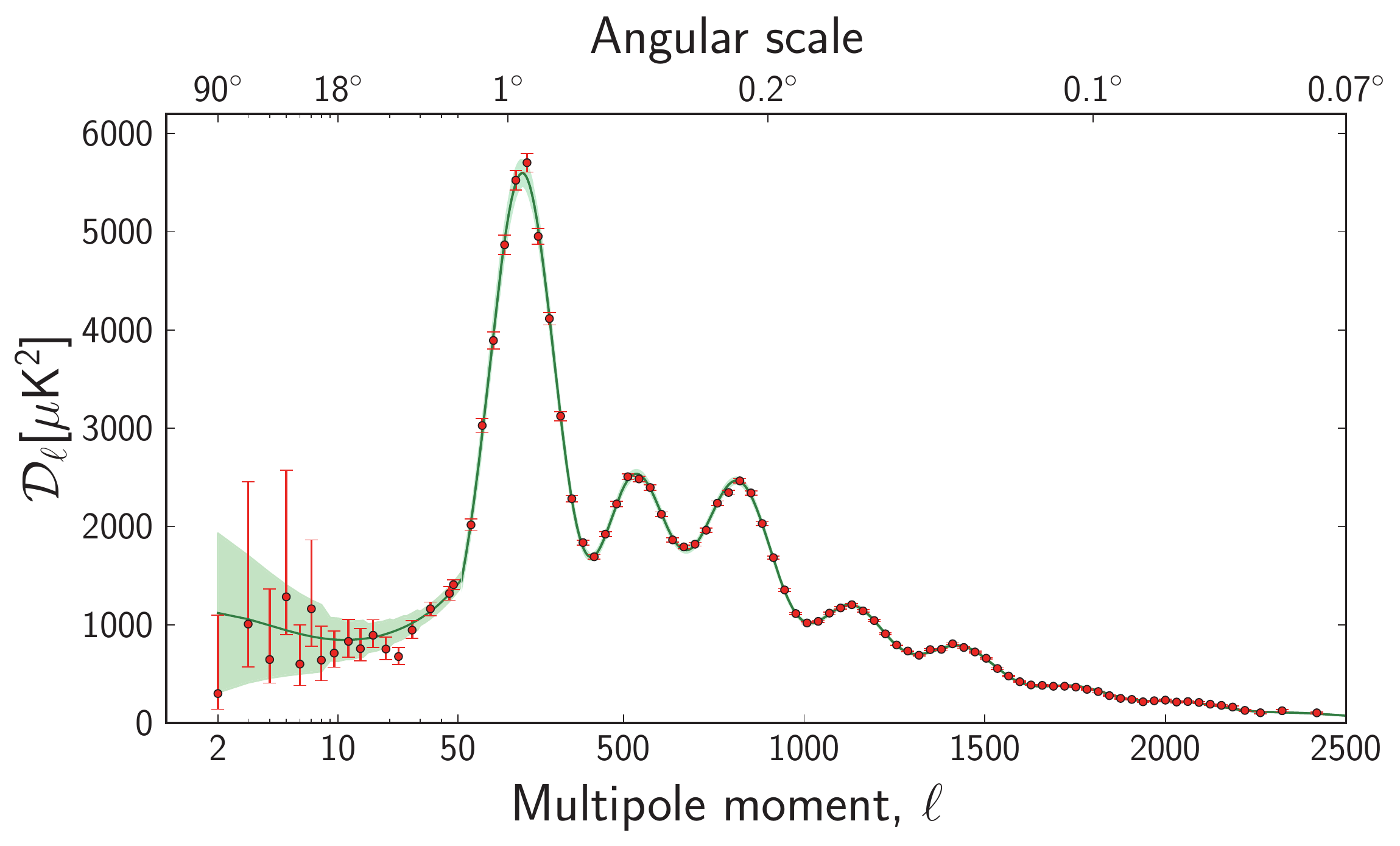}
\includegraphics[width=0.45\linewidth,clip=]{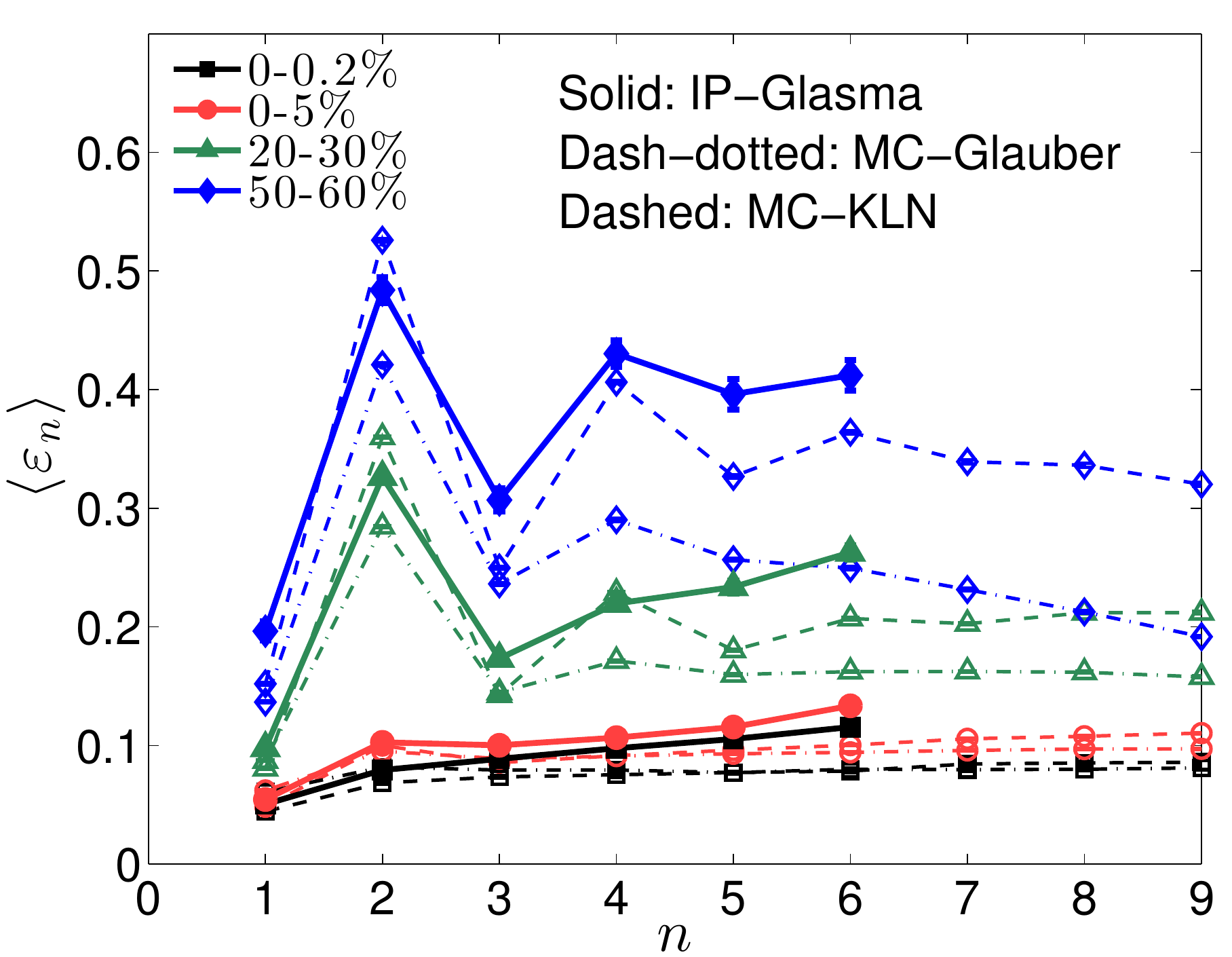}
\includegraphics[width=0.37\linewidth,clip=]{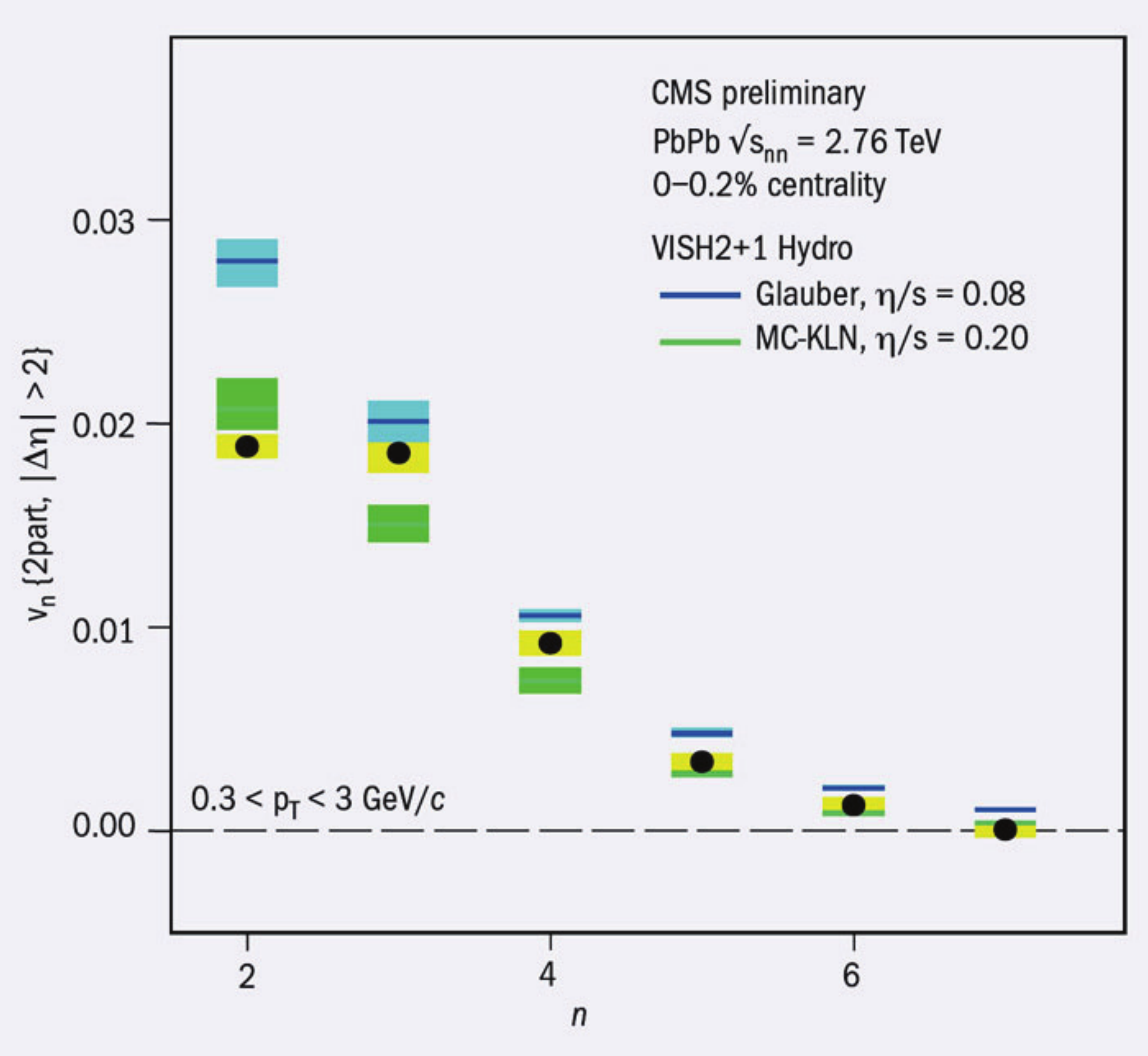}
\end{center}
\begin{spacing}{0.8}
\vspace*{-3mm}
\caption{
\label{F3}
{\footnotesize
{\sl Top:} Temperature power spectrum of the Big Bang at age 380,000 yr, as measured by the Planck satellite \cite{Planck:2013kta}.
{\sl Bottom left:} Primordial eccentricity power spectrum of the Little Bangs created in 2.76\,$A$\,TeV Pb+Pb collisions of different centralities, at $\tau{\,=\,}0$, from the energy density distributions of three different initial-state models (IP-Glasma, MC-Glauber, MC-KLN)
\cite{Heinz:2013th}. 
{\sl Bottom right:} The final Little Bang flow power spectrum for ultracentral (0-0.2\% centrality) Pb+Pb collisions at $\sqrt{s}{\,=\,}2.76\,A$\,TeV at the LHC, measured by the CMS Collaboration
(Wei Li, {\sl Quark Matter 2012}) and calculated with viscous hydrodynamics {\tt VISH2{+}1}
\cite{Shen:2011eg} using MC-Glauber and MC-KLN initial conditions with the indicated specific shear viscosities. Figure taken from \cite{CC}.
}
}
\end{spacing}
\end{figure}
%
between the fireball center and the surrounding vacuum, these fireballs undergo explosive collective expansion, cooling down rapidly through several different states of matter, finally fragmenting into thousands of free-streaming hadrons whose energy and momentum distributions can be detected in the detectors set up around the collider rings. The evolution history of these ``Little Bangs'' has much similarity with the Big Bang that created our Universe
(Fig.~\ref{F1}): Both undergo Hubble-like expansion,\footnote{The relative velocity between 
   	two matter elements increases roughly linearly with their relative distance.} 
feature a hierarchy of decoupling processes that are driven by the expansion dynamics (and not by the finite geometric size of the Little Bang), with chemical decoupling of the finally observed particle abundances\footnote{In the Big Bang the chemical composition is frozen during 
  	the process of primordial nucleosynthesis at an age of about 3 minutes, in the Little 
	Bang this happens during hadrosynthesis at the quark-hadron
	transition at an age of about 20-30 ioctoseconds \cite{Heinz:1998nk}.}
preceding kinetic decoupling,\footnote{In the Big Bang, the formation of neutral atoms by charge
	recombination at an age of about 380,000 yr makes the universe transparent to light and 
	freezes the thermal Bose-Einstein energy distribution of the Cosmic Microwave 
	Background radiation. When the 
	Little Bang reaches an age of about 40-50 ioctoseconds, the final-stage hadron gas
	becomes so dilute that strong interactions between the hadrons cease and their 
	energies and momenta are ``frozen out''.} 
and with initial-state quantum fluctuations\footnote{The initial wave function of the universe,
	stretched by cosmic inflation, seeds density fluctuations in the Big Bang. In the Little Bang,
	the initial quark and gluon wave functions inside the nucleons within the colliding nuclei 
	control its initial density distribution.} 
(shown in Figures~\ref{F2} and \ref{F3}) imprinting themselves\footnote{In the Big Bang
  	the initial density fluctuations evolve under the action of gravity, described by Einstein's
	general theory of relativity, into the observed Cosmic Microwave temperature fluctuation
	spectrum shown in Figs.~\ref{F2} (left) and \ref{F3} (top), and ultimately into 
	today's distributions of stars, galaxies, and galaxy clusters and superclusters (see top 
	panel of Fig.~\ref{F1}). In the Little Bang, they evolve through viscous hydrodynamics 
	into the measured anisotropic collective flow patterns and their event-by-event 
	fluctuations that are the subject of this overview.}
onto the experimentally observed final state.
 
Of course, the Big and Little Bangs are quite different in other aspects: Their expansion rates differ by about 18 orders of magnitude; the Little Bang's expansion is 3-dimensional and driven by pressure gradients, not 4-dimensional and controlled by gravity; Little Bangs evolve on time scales of ioctoseconds, not billions of years; distances are measured in femtometers rather than light years. Most importantly, the Little Bang Standard Model is still under construction. This overview discusses recent progress of the edifice. 
 
\section{Eccentricity fluctuations, anisotropic flows, and flow fluctuations}
\label{sec2}
  
\begin{figure}[htb]
\vspace*{-2mm}
\includegraphics[width=0.32\linewidth,clip=]{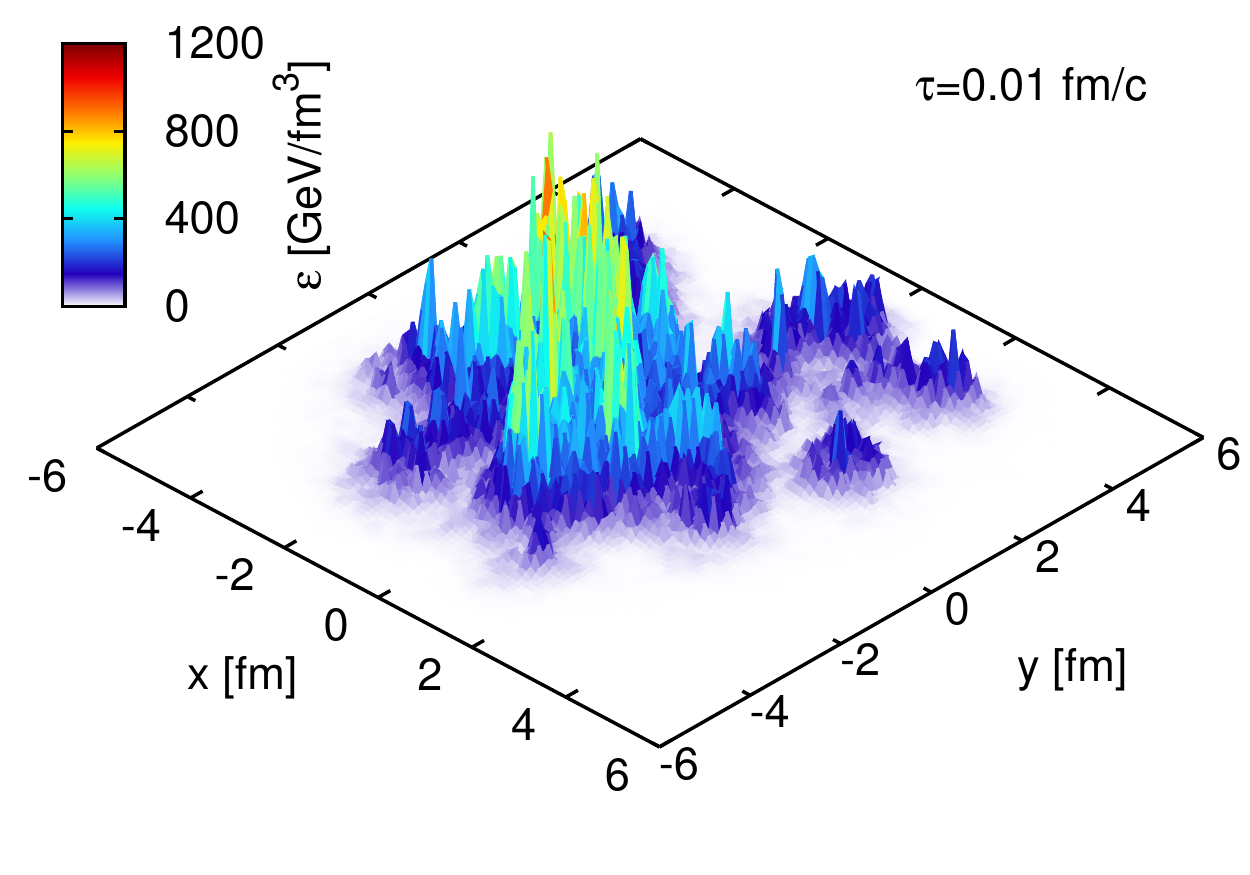}
\includegraphics[width=0.32\linewidth,clip=]{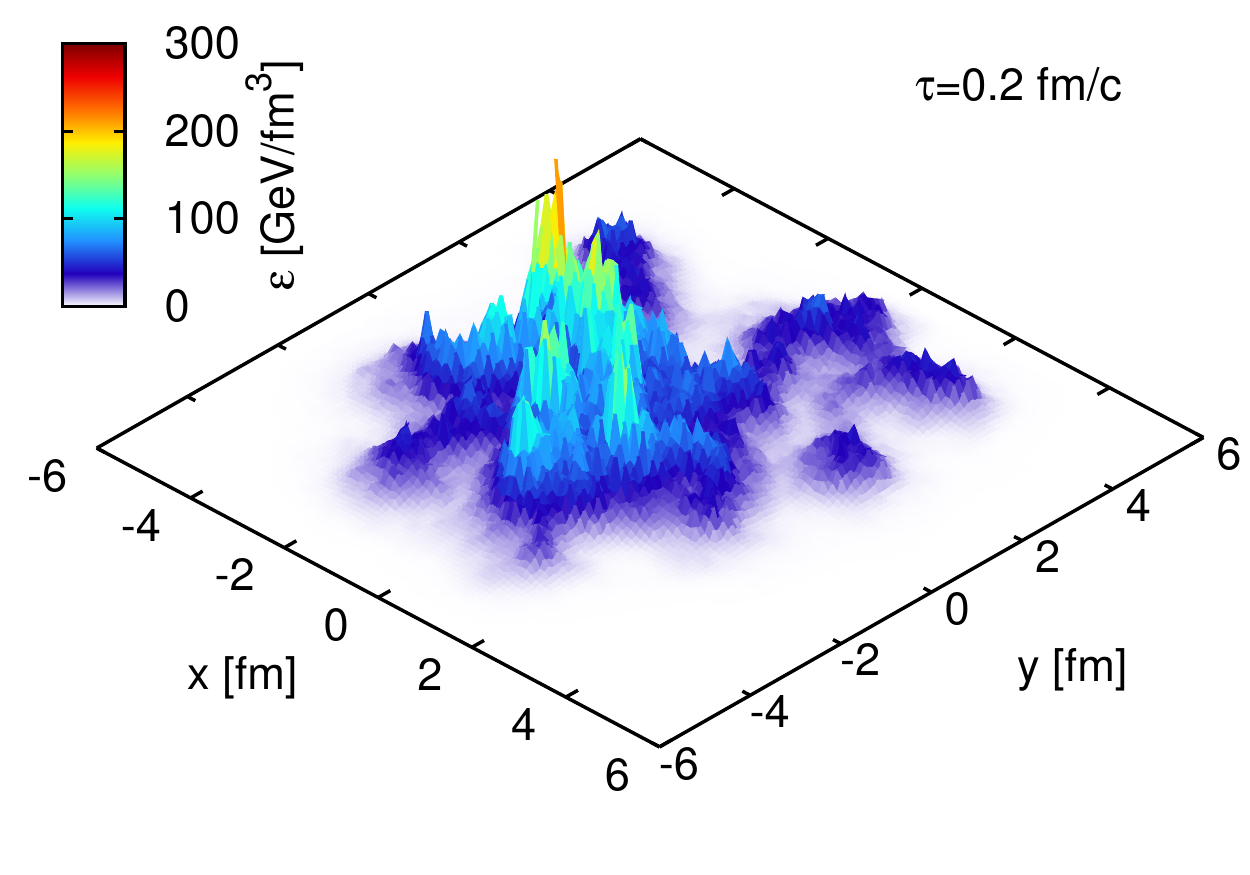}
\includegraphics[width=0.32\linewidth,clip=]{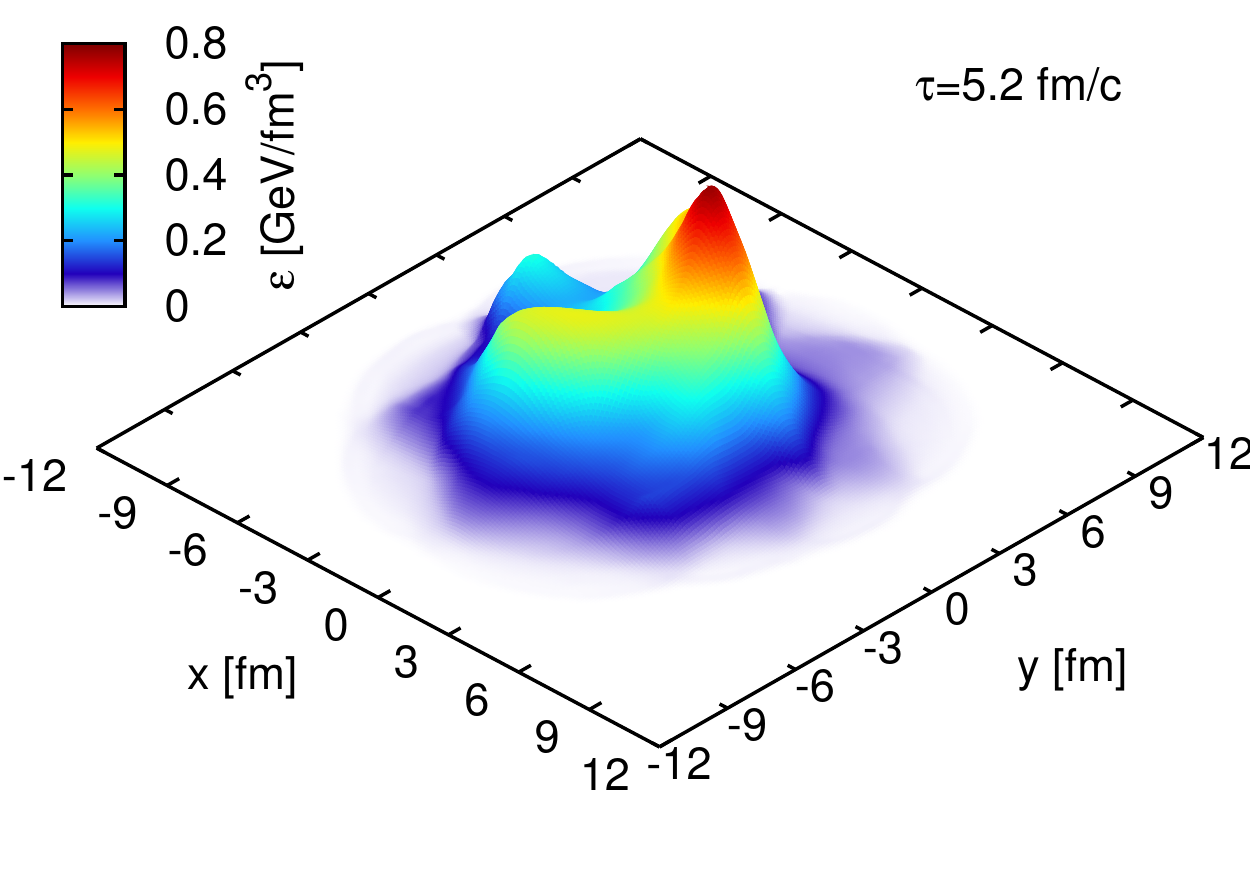}
\begin{spacing}{0.8}
\vspace*{-5mm}
\caption{
\label{F4}
{\footnotesize
Typical transverse energy density profiles $e(x,y)$ from the IP-Glasma model \cite{Schenke:2012wb} for a semiperipheral ($b{\,=\,}8$\,fm) Au+Au collision at $\sqrt{s}{\,=\,}200\,A$\,GeV, at times $\tau=0.01,\ 0.2,$ and 5.2\,fm/$c$. From $\tau{\,=\,}0.01$\,fm/$c$ to 0.2\,fm/$c$ the fireball evolves out of equilibrium according to the Glasma model \cite{Kovner:1995ja}; at $\tau{\,=\,}0.2$\,fm/$c$ the energy momentum tensor from the IP-Glasma evolution is Landau-matched to ideal fluid form (for technical reasons \cite{Gale:2012rq} the viscous pressure components are set to zero at the matching time) and henceforth evolved with viscous Israel-Stewart fluid dynamics, assuming $\eta/s{\,=\,}0.12$ for the specific shear viscosity. The pre-equilibrium Glasma evolution is seen to somewhat wash out the large initial energy density fluctuations. The subsequent dissipative hydrodynamic evolution further smoothes these fluctuations. The asymmetric pressure gradients due to the prominent dipole asymmetry in the initial state of this particular event (visible as a left-right asymmetry of the density profile in the left panel) is seen to generate a dipole (``directed flow'') component in the hydrodynamic flow pattern that pushes matter towards the right during the later evolution stages.
}
\vspace*{-2mm}
}
\end{spacing}
\end{figure}

We can observe only one Big Bang (the one that produced our universe), but at the Relativistic Heavy-Ion Collider (RHIC) and Large Hadron Collider (LHC) we have experimentally created and studied billions of Little Bangs. Each Little Bang is different: Highly successful phenomenology based on hydrodynamic evolution models \cite{Kolb:2003dz,Heinz:2013th} has taught us that the initially very dense quark-gluon matter created in heavy-ion collisions reaches approximate local thermal equilibrium on a very short time scale \cite{Heinz:2001xi} of order 1\,fm/$c$ (3 ioctoseconds), after which it evolves according to the macroscopic laws of relativistic viscous fluid dynamics. Pressure gradients in the fluctuating density profile (see Fig.~\ref{F4} for a specific example) are the hydrodynamic forces that accelerate the fluid and cause it to expand and dilute. Spatial anisotropies and inhomogeneities in the initial density profile transverse to the beam direction lead to corresponding anisotropies in the final transverse expansion flow velocity profile that are imprinted on the momenta of the experimentally observed particles emitted from the collision. Each Little Bang features its own final flow velocity profile which (within hydrodynamics) is a deterministic classical response to the initial conditions that fluctuate from collision to collision due to quantum fluctuations in the initial nuclear wave function. We now know that the quark-gluon liquid that makes up the matter of the Little Bang during the first half of its life has very small viscosity, behaving like an almost ideal fluid \cite{Kolb:2003dz}. This is a fantastic gift of Nature since it allows us to study experimentally the spectrum of initial-state quantum fluctuations through the final-state anisotropic flow fluctuations. Had the quark-gluon plasma (QGP) turned out to be highly viscous, all initial-state fluctuations would have been wiped out by dissipation before final decoupling of the emitted particles, thereby closing this observation window on the initial state of the Little Bang and on the quantum nature of the initial energy deposition process.   

The Little Bang pressure $p$ is related to its energy density $e$ by its equation of state (EOS), and the anisotropy of the initial pressure gradients can be characterized by a series of harmonic eccentricity coefficients 
\begin{equation}
\label{eq1}
\varepsilon_1 e^{i\Phi_1} \equiv - \frac{\int  r \, dr \, d\varphi \, r^3 e^{i\varphi} \, e(r,\varphi)}
                                                            {\int r \, dr \, d\varphi \, r^3 e(r,\varphi)}, \qquad
\varepsilon_n e^{in\Phi_n} \equiv - \frac{\int  r \, dr \, d\varphi \, r^n e^{in\varphi} \, e(r,\varphi)}
                                                              {\int r \, dr \, d\varphi \, r^n e(r,\varphi)}\  (n>1),
\end{equation}
where $e(r,\varphi)$ is the initial energy density distribution in the plane transverse to the beam direction. $\varepsilon_n$ characterizes the magnitude of the $n$th harmonic deformation coefficient, and the angle $\Psi_n$ gives the direction of the corresponding deformation component in the lab frame. Both the magnitudes $\varepsilon_n$ and their orientations $\Psi_n$ fluctuate from event to event. The mean values $\langle \varepsilon_n\rangle$ averaged over many events from three popular initial energy deposition models (see \cite{Heinz:2013th} for a description and original references) are shown as a function of harmonic index $n$ in the left bottom panel of Fig.~\ref{F3}. They represent the primordial temperature fluctuation spectrum of the Little Bang. We see that each collision centrality generates a different class of Little Bangs, with its own $\langle \varepsilon_n\rangle$ power spectrum, and that the three models shown give quite different results for these power spectra. If we can somehow measure the centrality dependence of the initial $\langle \varepsilon_n\rangle$ power spectrum, we have a powerful constraint on the initial nuclear quark and gluon wave functions on our hands. This is reminiscent of the crucial role the CMB temperature power spectrum (shown in the top panel of Fig.~\ref{F3}) has played in nailing down the parameters of the Standard Model of Big Bang Cosmology.

The hydrodynamic response to the initial $\varepsilon_n$-spectrum of the Little Bang is reflected in the observed transverse momentum distributions $dN_i/(dyp_Tdp_Td\phi)$ of the various emitted hadronic species $i$, and in correlations among them. The transverse momentum spectra can again be characterized by a set of complex harmonic coefficients 
\begin{eqnarray} 
  \label{eq2}
  V_n &=& v_n e^{in\Psi_n} 
  := \frac{\int \pT d\pT d\phi\, e^{i n \phi} \,\frac{dN}{dy \pT d\pT d\phi}}
                  {\int \pT d\pT d\phi\, \frac{dN}{dy \pT d\pT d\phi}}
   \equiv \{e^{in\phi}\},
  \\
\label{eq3}
   V_n(\pT) &=& v_n(\pT) e^{in\Psi_n(\pT)} 
   := \frac{\int d\phi \, e^{i n \phi}\, \frac{dN}{dy \pT d\pT d\phi}} 
                   {\int d\phi \, \frac{dN}{dy \pT d\pT d\phi}}
   \equiv \{e^{in\phi}\}_{\pT}.
\end{eqnarray}
Here $\phi$ is the azimuthal angle around the beam direction of the particle's transverse momentum $\bm{p}_\mathrm{T}$, and the curly brackets denote the average over particles from a single collision. Each particle species $i$ has its own set of anisotropic flow coefficients $V_n$. Eq.\,(\ref{eq2}) defines the flow coefficients and associated flow angles for the entire event, whereas Eq.\,(\ref{eq3}) is the analogous definition for the subset of particles in the event with a given magnitude of the transverse momentum $\pT$. I suppress the dependence of both types of flow coefficients on the rapidity $y$. $v_n$ are known as the ``integrated'' anisotropic flows, $v_n(\pT)$ are called ``differential'' flows. By definition, both $v_n$ and $v_n(\pT)$ are positive definite. Hydrodynamic simulations show that in general the flow angles $\Psi_n$ depend on $\pT$, and that, as a function of $\pT$, $\Psi_n(\pT)$ wanders around the ``average angle'' $\Psi_n$ that characterizes the integrated flow $v_n$ of the entire event \cite{Heinz:2013bua}.

For each collision system, centrality class, and collision energy, the fluctuating initial state of the Little Bang is characterized by a distinct probability distribution $P(\varepsilon_n,\Phi_n)$, one particular moment of which is the $\langle \varepsilon_n\rangle$ power spectrum shown in Fig.~\ref{F3}. For different $n$, the angles $\Phi_n$ are more or less correlated with the direction of the impact parameter $\bm{b}$, but this direction cannot be measured experimentally. The viscous hydrodynamic evolution relates the probability distribution for the initial complex eccentricity coefficients deterministically to probability distributions $P(V_n)$ and $P(V_n(\pT))$ for the integrated and differential final complex harmonic flow coefficients; these distributions are particle species specific. As will be discussed in the next section, the relation between the final complex flow coefficients $V_n$ and the initial eccentricity coefficients $\varepsilon_n e^{in\Phi_n}$ (and thus between the corresponding flow and eccentricity probability distributions characterizing each class of collisions) depends on the viscosity of the Little Bang matter. One goal of the relativistic heavy-ion program is to both constrain the QGP viscosity and identify the correct theory for the initial-state quantum fluctuations by performing a complete experimental reconstruction of the final multi-dimensional distributions $P(V_n)$ and $P(V_n(\pT))$.  
  
Due to limited statistics arising from the finite number of particles emitted by each Little Bang, neither the magnitudes $v_n$ nor the flow angles $\Psi_n$ can be accurately determined for a single event. Experimental flow measures therefore involve angle correlations between two or more particles; for example, $\langle \{e^{in\phi_1}\}_{p_{T1}}\{e^{-in\phi_2}\}_{p_{T2}} \rangle = \langle v_n(p_{T1}) v_n(p_{T2}) \cos[n(\Psi_n(p_{T1}){-}\Psi_n(p_{T2}))]\rangle$ where the first particle from the event has transverse momentum $\bm{p}_{T1}=(p_{T1},\phi_1)$ and the second particle has $\bm{p}_{T2}=(p_{T2},\phi_2)$, the average $\{\dots\}$ is over all such particles in the event and $\langle\dots\rangle$ indicates the average of the result over many Little Bangs of the selected class. One sees that the event-by-event fluctuations and transverse momentum dependences of both the anisotropic flow magnitudes $v_n$ and their associated flow angles $\Psi_n$ affect these experimental observables. Different such observables correspond to different correlation functions between the $v_n$'s and $\Psi_n$'s all of which can, for a given initial energy deposition model with probability distribution $P(\varepsilon_n,\Phi_n)$, be computed from the hydrodynamically predicted probability distributions $P(V_n)$ and $P(V_n(\pT))$. One can define experimental observables that separate the fluctuations of the anisotropic flow magnitudes $v_n(\pT)$ from those in the flow angles $\Psi_n(\pT)$ \cite{Heinz:2013bua}. These should be powerful discriminators between different $P(V_n)$ and $P(V_n(\pT))$ distributions, and thus between different initial-state fluctuation models. Experimental studies of these $v_n$ and $\Psi_n$ fluctuations and their $\pT$-dependences have just gotten under way; the results will be interesting and should significantly advance the construction of the Little Bang Standard Model.

\begin{figure}[htb]
\vspace*{-2mm}
\includegraphics[width=0.49\linewidth,clip=]{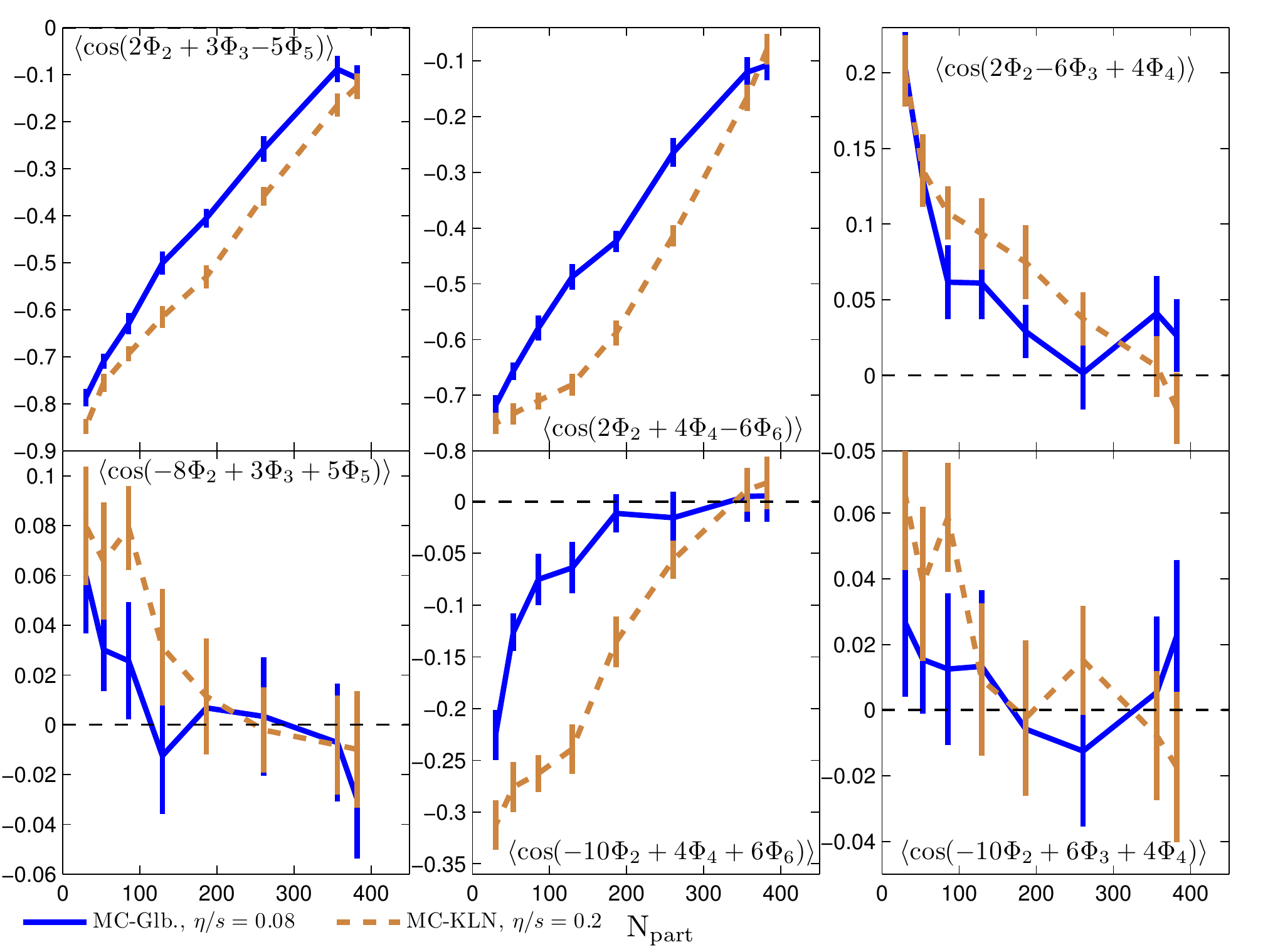}
\includegraphics[width=0.49\linewidth,clip=]{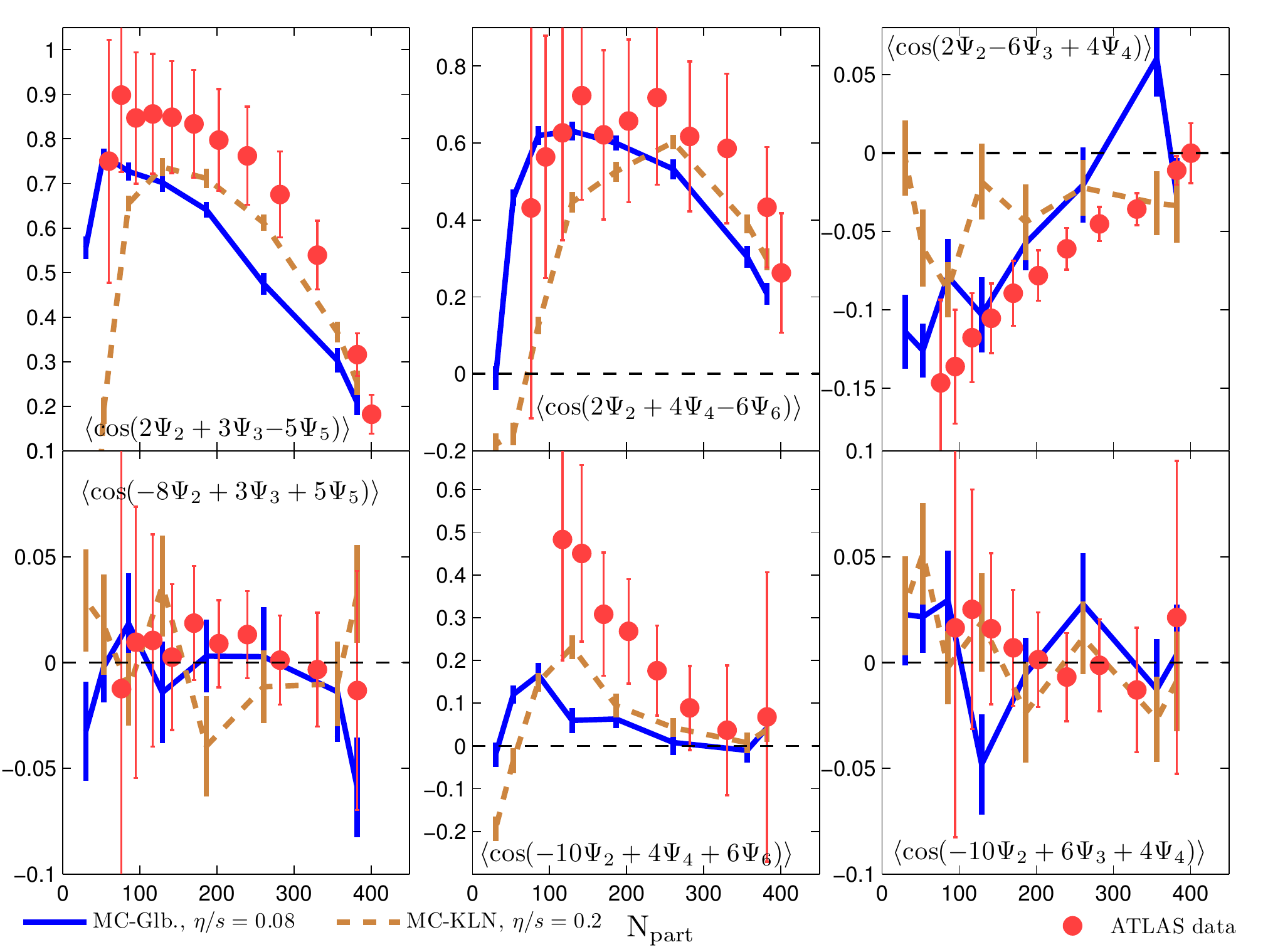}
\begin{spacing}{0.8}
\vspace*{-2mm}
\caption{
\label{F5}
{\footnotesize
Three-plane correlations between the initial participant planes (left 6-panel) and the final-state flow planes (right 6-panel), for $2.76\,A$\,TeV Pb+Pb collisions as functions of collision centrality, indicated by the number of participant nucleons $N_\mathrm{part}$ on the horizontal axes. The solid (dashed) lines are for MC-Glauber (MC-KLN) initial conditions evolved event-by-event with viscous hydrodynamics using $\eta/s{\,=\,}0.08$ (0.2) for the specific shear viscosity
\cite{Qiu:2012uy}. Filled circles show the experimental values measured by the ATLAS Collaboration \cite{Jia:2012sa}.
}
\vspace*{-1mm}
}
\end{spacing}
\end{figure}

The angles $\Phi_n$ associated with the initial deformation parameters $\varepsilon_n$ are not only correlated with the (unmeasurable) direction of the impact parameter $\bm{b}$, but also with each other, in ways that are predicted by (and thus depend on) the initial energy deposition model \cite{Jia:2012ma,Qiu:2012uy}. These so-called participant-plane correlations in the initial state are translated by hydrodynamic response into final-state flow angle correlations. These flow angle correlations were measured in Pb+Pb collisions at the LHC by the ALICE \cite{ALICE:2011ab} and ATLAS \cite{Jia:2012sa} Collaborations. All measured two- and three-plane flow angle correlations \cite{Jia:2012sa} are qualitatively reproduced by viscous hydrodynamic calculations, both in their magnitudes and centrality dependences \cite{Qiu:2012uy} (right panels in Fig.\ \ref{F5}). They differ, however, qualitatively from the initial participant-plane correlations \cite{Qiu:2012uy}, often even in sign (left panels in Fig.\ \ref{F5}). Hydrodynamic evolution is nonlinear and leads to mode-coupling between different harmonics \cite{Qiu:2011iv}. For example, elliptic and triangular deformations $\varepsilon_2$ and $\varepsilon_3$ provide a nonlinear contribution to pentangular flow $V_5$ which, for large impact parameters where the elliptic deformation $\varepsilon_2$ is big, overwhelms the linear response to $\varepsilon_5$ and completely decorrelates the pentangular flow plane $\Psi_5$ from the angle $\Phi_5$ of the initial pentangular density deformation \cite{Qiu:2011iv}. Only by accounting for mode-coupling, either directly by following the non-linear hydrodynamic evolution \cite{Qiu:2012uy} or through a non-linear response analysis that keeps at least second-order terms \cite{Teaney:2012ke}, can the experimental data be reproduced. These flow-plane correlations thus represent an {\em experimentum crucis} in support of the hydrodynamic paradigm for the Little Bang; dynamical models without a large degree of local thermalization and hydrodynamic collective flow will not be able to describe the dynamical change of character between the initial-state participant-plane and final-state flow-plane correlations \cite{Qiu:2012uy}.\footnote{The strength of the flow-plane correlations depends on the QGP shear viscosity \cite{Qiu:2012uy}. A careful quantitative study of these correlations 
    	by more precise experiments and more systematic theoretical analyses should be able 
	to separate the effects arising from the initial-state fluctuation spectrum and from 
	dissipative transport.} 
 
\section{QGP shear viscosity from anisotropic flow measurements}
\label{sec3}

\begin{figure}[htb]
\includegraphics[width=0.63\linewidth,clip=]{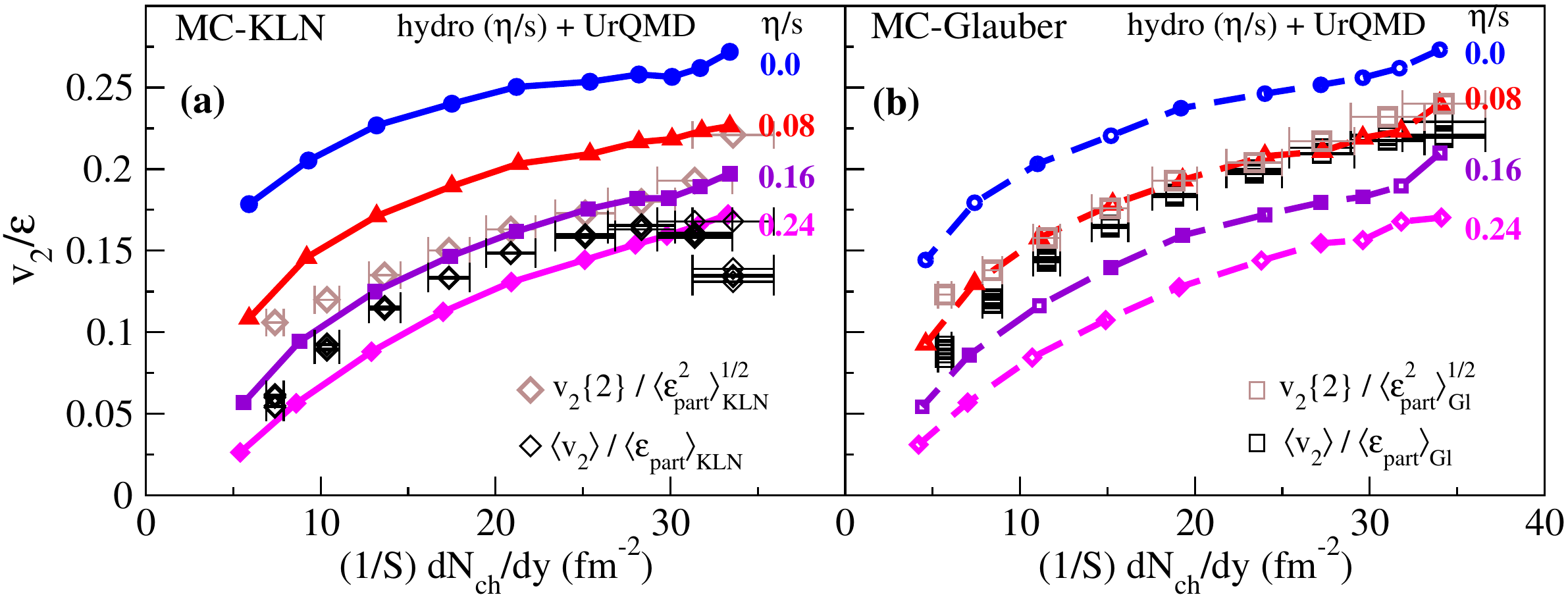}
\includegraphics[width=0.365\linewidth,clip=]{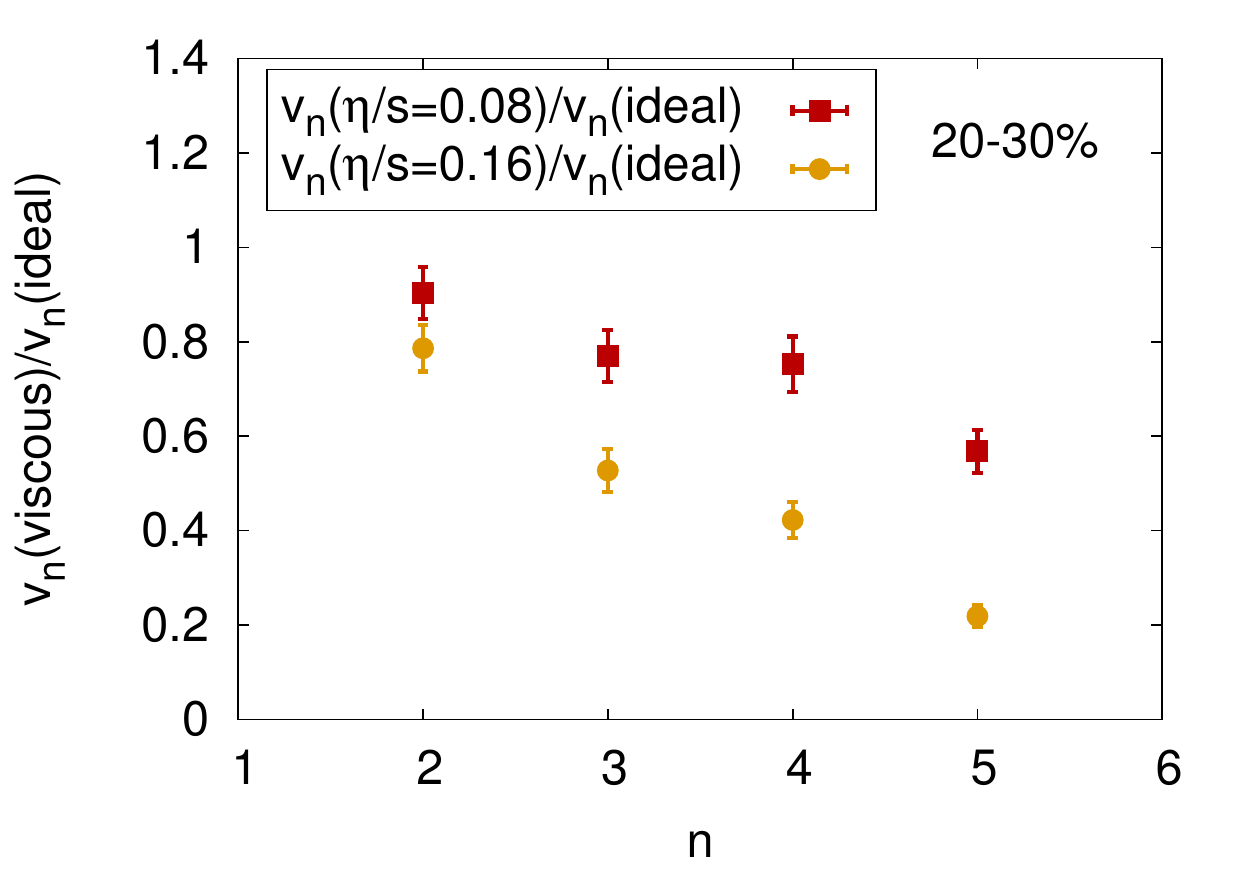}
\begin{spacing}{0.8}
\vspace*{-4mm}
\caption{
\label{F6}
{\footnotesize
{\sl Left:} The eccentricity-scaled integrated elliptic flow of all charged hadrons, $v_2^\mathrm{ch}(\eta/s)/\varepsilon$, as a function of total charged hadron multiplicity density per unit overlap area, $(1/S)(dN_\mathrm{ch}/dy)$. The experimental data points show two measures for the elliptic flow ($\langle v_2\rangle$ \cite{Ollitrault:2009ie} and $v_2\{2\}$ \cite{Adams:2004bi}) from 200\,$A$\,GeV Au-Au collisions at RHIC, measured by the STAR Collaboration. Both panels use the same sets of data, but use different average initial eccentricities $\langle \varepsilon\rangle$ and overlap areas $\langle S\rangle$ (obtained from the MC-Glauber and MC-KLN models) to normalize the vertical and horizontal axes. The theoretical curves were computed with the {\tt VISHNU} model \cite{Song:2010aq}, for different (temperature-independent) choices of the specific QGP shear viscosity $(\eta/s)_\mathrm{QGP}$.
{\sl Right:} Viscous suppression of $v_n$ for $\eta/s{\,=\,}0.08$ (squares) and 0.16 (circles), as function of harmonic index $n$, for Au+Au collisions at 20-30\% centrality \cite{Schenke:2011bn}.
}
}
\end{spacing}
\end{figure}

The efficiency of the fluid to convert spatial anisotropies in the pressure gradients into anisotropic flows is degraded by shear viscosity. The ``conversion efficiency'' $v_n/\varepsilon_n$ is therefore a measure for the specific shear viscosity $\eta/s$ of the expanding fluid. This is shown in Fig.~\ref{F6}. Higher harmonics, reflecting variations on smaller spatial scales, are suppressed more strongly by shear viscosity than lower flow harmonics.
%
\vspace*{-4mm}
%

The azimuthally symmetric part of the collective transverse flow, called radial flow, boosts particles from low to high transverse momentum in proportion to their masses; it is thus responsible for the distribution of the hydrodynamically generated momentum anisotropies over the various particle species and in $\pT$ \cite{Song:2011hk}. To correctly describe the differential anisotropic flows $v_n(\pT)$, for all charged hadrons or for specific identified hadron species, thus requires that the total hydrodynamic momentum anisotropy (which ``measures'' $\eta/s$), the radial flow (which is also sensitive to bulk viscosity that suppresses radial flow \cite{Song:2009rh}), and the chemical composition of the system at final decoupling (which reflects the kinetics of chemical freeze-out) are {\em all} correctly described by the model. By analyzing instead the total flow anisotropies $v_n^\mathrm{ch}$ of {\em all} charged hadrons, integrated over $\pT$, one can strongly reduce the model sensitivity to bulk viscosity and final chemical composition and obtain a much more robust estimate of $\eta/s$ \cite{Song:2010aq,Song:2011hk}. The left panel in Fig.~\ref{F6} shows such an extraction of $\eta/s$ from elliptic flow data. The {\tt VISHNU} model used for this extraction couples the viscous hydrodynamic evolution of the QGP phase to a microscopic kinetic evolution of the hadronic phase after hadronization, thereby eliminating many uncertainties in earlier simplified calculations that described the hadronic phase macroscopically, too. The different theoretical lines show that, for a given initial energy deposition model, $v_2^\mathrm{ch}/\varepsilon_2$ is {\em only} sensitive to the specific QGP shear viscosity. Once $(\eta/s)_\mathrm{QGP}$ has been adjusted to the measured total charged hadron elliptic flow as in Fig.~\ref{F6} (resulting in $(\eta/s)_\mathrm{QGP}{\,\simeq\,}0.2$ for MC-KLN and ($\eta/s)_\mathrm{QGP}{\,\simeq\,}0.08$ for MC-Glauber initial conditions), one finds that the model also correctly describes the $\pT$-spectra and differential elliptic flows $v_2(\pT)$ of all charged hadrons together, as well as for specific identified hadron species (pions, kaons, protons), for all collision centralities \cite{Song:2011hk}.
Furthermore, it correctly {\em predicted} the analogous observables for Pb+Pb collisions at the LHC \cite{Shen:2011eg,Heinz:2011kt,Abelev:2012wca}, in particular an increased mass splitting between the differential elliptic flows for light and heavy hadrons \cite{Shen:2011eg,Muller:2012zq}.

\begin{figure}[htb]
\vspace*{-3mm}
\begin{center}
\begin{minipage}{8.2cm}
\includegraphics[width=8.2cm,clip=]{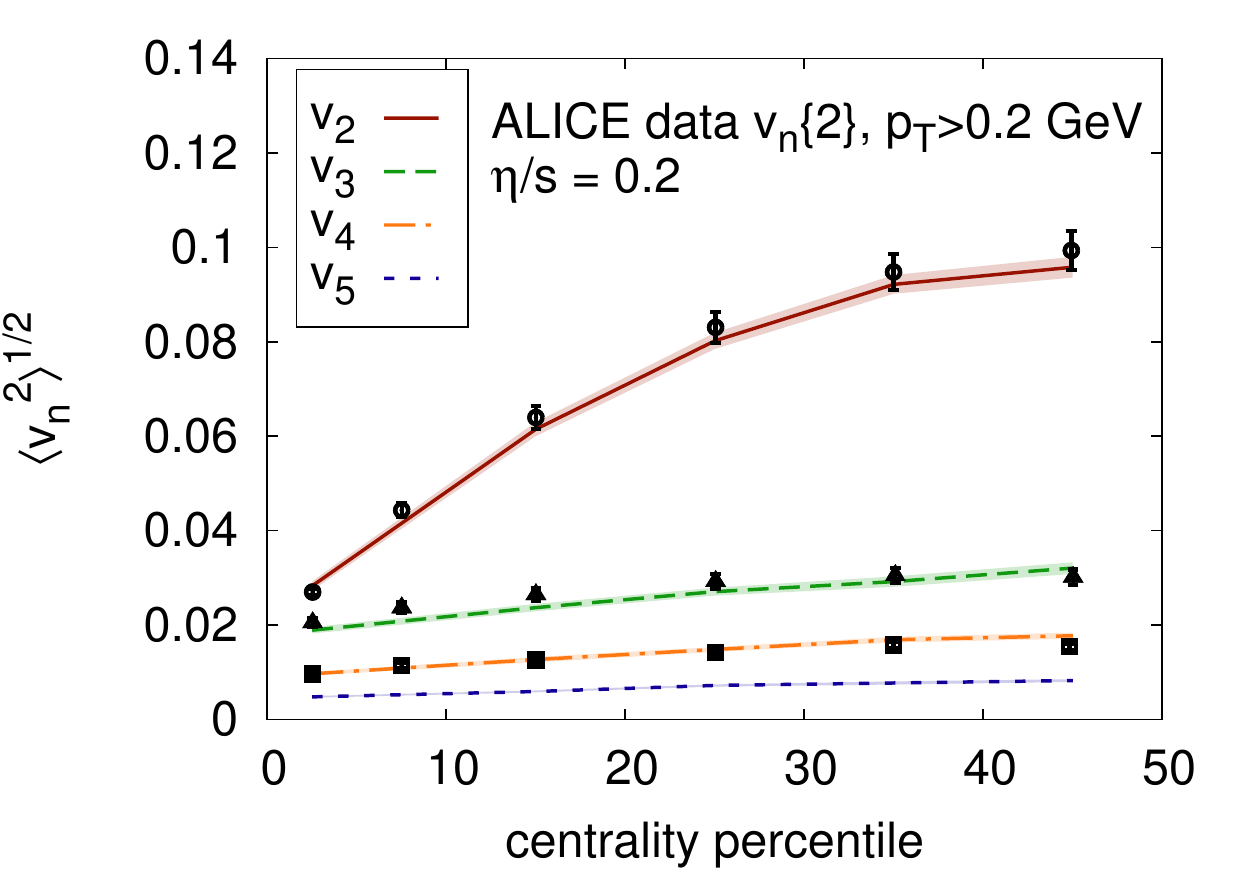}\\
\includegraphics[width=8.2cm,clip=]{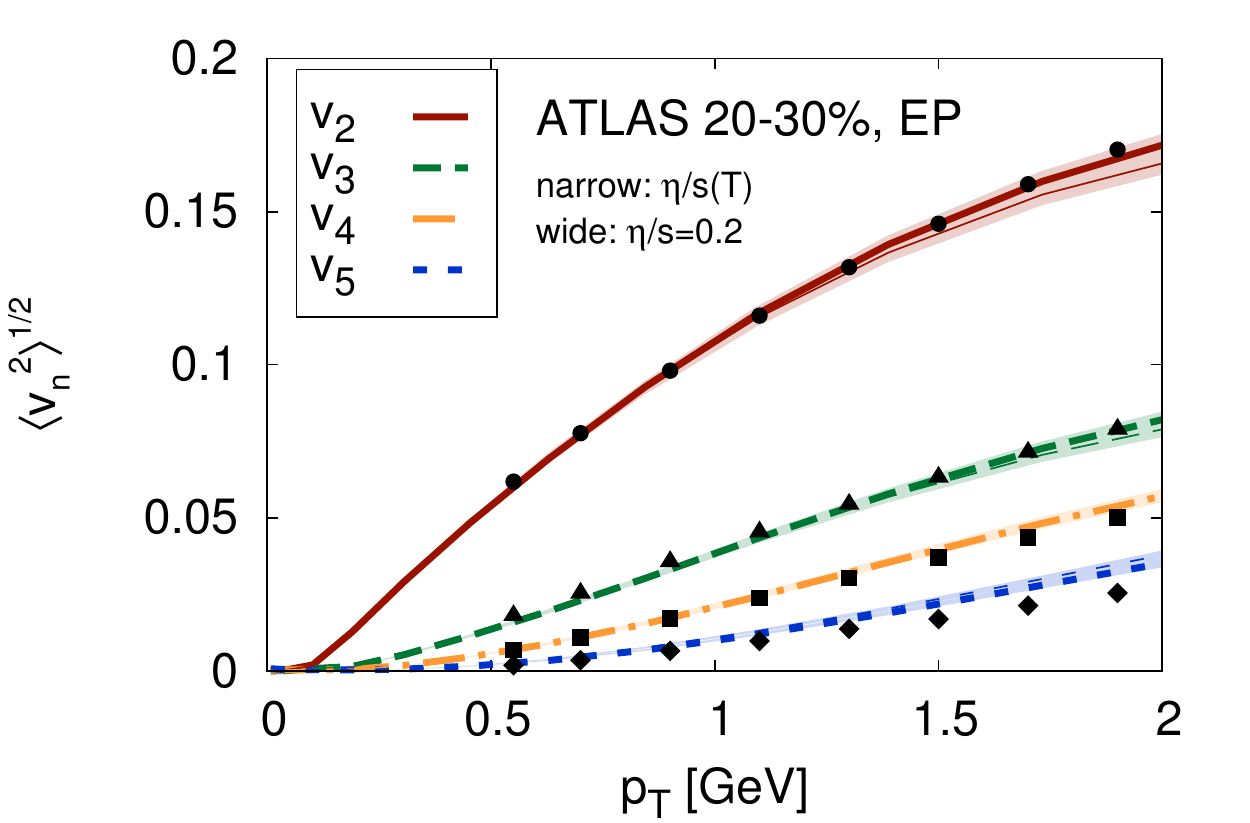}
\end{minipage}
\begin{minipage}{7cm}
\includegraphics[width=7cm,clip=]{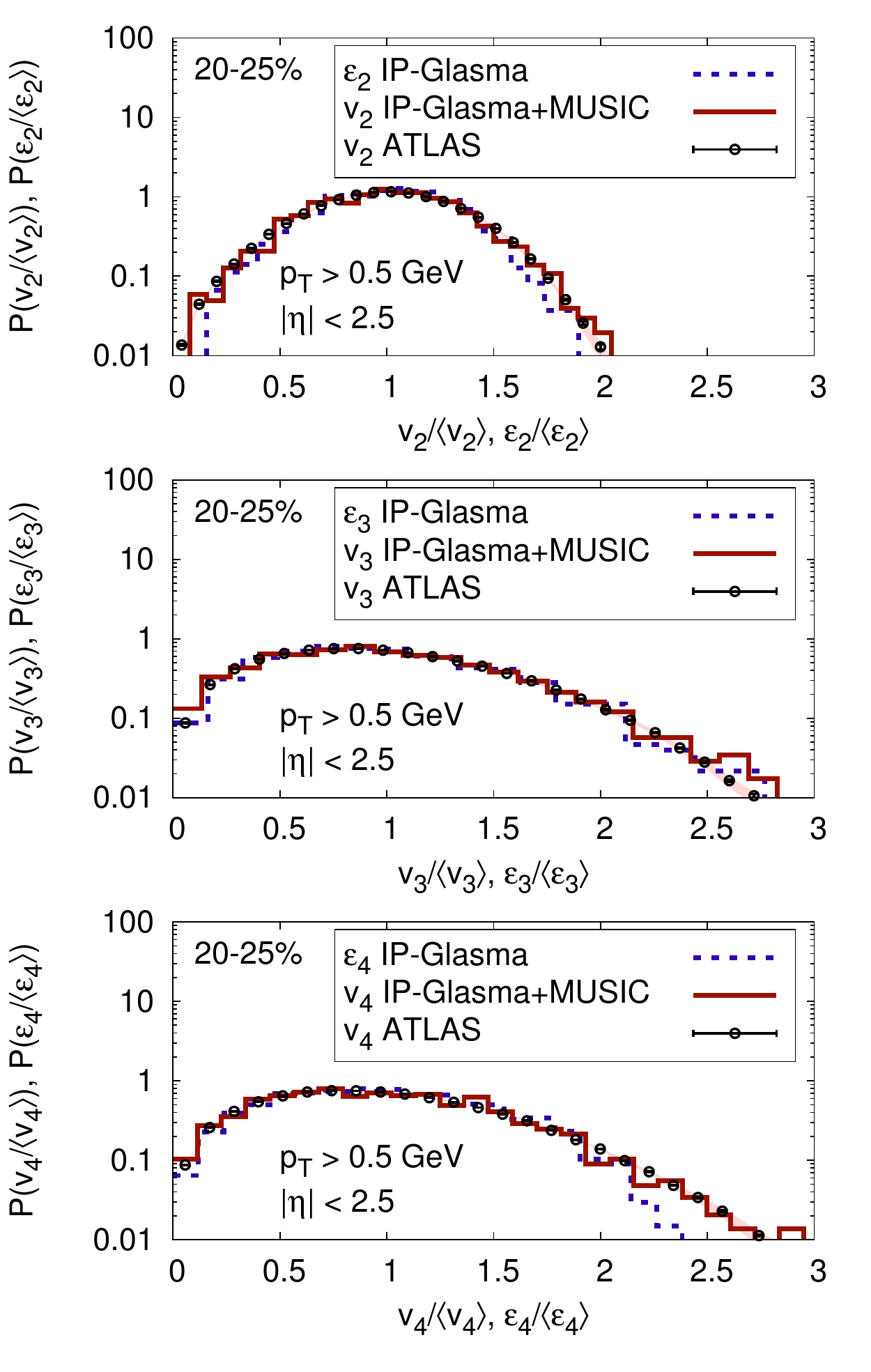}
\end{minipage}
\end{center}
\vspace*{-5mm}
\begin{spacing}{0.8}
\caption{
\label{F7}
{\footnotesize
{\sl Top left:} The centrality dependence of $v_n\{2\}$ from 2.76\,$A$\,TeV Pb+Pb collisions measured by ALICE \cite{ALICE:2011ab} compared to viscous hydrodynamic model calculations \cite{Gale:2012rq}.
{\sl Bottom left:} Comparison of $v_n(\pT)$ for the same collision system at $20{-}30\%$ centrality from ATLAS \cite{ATLAS:2012at} with hydrodynamical calculations, using both a constant average and a temperature dependent $\eta/s$ \cite{Gale:2012rq}.
{\sl Right:}
Scaled distributions of $v_{2,3,4}$ (from top to bottom) from viscous hydrodynamics with IP-Glasma initial conditions \cite{Gale:2012rq} compared with experimental data from ATLAS~\cite{Jia:2012ve} and with the scaled distributions of the corresponding initial eccentricities $\varepsilon_{2,3,4}$. Nonlinear hydrodynamic evolution causes slightly larger variances for the $v_n$ distributions compared to those of $\varepsilon_n$. The data in both panels are from Pb+Pb collisions at the LHC.
}
\vspace*{-2mm}
}
\end{spacing}
\end{figure}

Unfortunately, the two left panels in Fig.~\ref{F6} also show that different initial energy deposition models lead to very different estimates for $(\eta/s)_\mathrm{QGP}$, due to their different initial eccentricities $\varepsilon_2$ (see Fig.~\ref{F3}). In \cite{Qiu:2011hf} it was shown that this model ambiguity can be resolved by analyzing simultaneously elliptic and triangular flow ($v_2$ and $v_3$) data. This analysis eliminated the MC-KLN model with $(\eta/s)_\mathrm{QGP}{\,=\,}0.2$ as a viable candidate, and thereby demonstrated the power of a comprehensive set of anisotropic flow data to (over)constrain the dynamical evolution model.  Unfortunately, the MC-Glauber model did not enjoy a much longer life itself: The $v_n$ data ($n{\,=\,}2,\dots,7$) from ultra-central Pb+Pb collisions shown by CMS at {\sl Quark Matter 2012} (see the right bottom plot in Fig.~\ref{F3}) and the event-by-event $v_n$ probability distributions ($n{\,=\,}2,\dots,4$) measured by ATLAS and shown at the same conference \cite{Jia:2012ve} (see the right panel in Fig.~\ref{F7}) can not simultaneously be described by viscous hydrodynamics for {\em any} choice of $(\eta/s)_\mathrm{QGP}$ if initial fluctuation spectra from either of these two models (MC-KLN and MC-Glauber) are used. 

Fortunately, a new and much more successful initial-state model saved the day: the IP-Glasma model \cite{Schenke:2012hg,Schenke:2012wb}, based on the Color Glass Condensate idea \cite{Kovner:1995ja}, implements gluonic field fluctuations inside nucleons \cite{Tribedy:2010ab,Schenke:2012wb} as well as gluon saturation effects \cite{Kovner:1995ja,Bartels:2002cj}. As shown in Fig.~\ref{F7}, these IP-Glasma initial conditions (represented by the solid lines in Fig.~\ref{F3}, lower left panel), when (after a short initial pre-equilibrium stage modeled by classical Yang-Mills evolution) evolved with viscous fluid dynamics, reproduce the entire measured spectrum of charged hadron anisotropic flow coefficients $v_n$, both integrated over and differential in $\pT$, for all collision centralities, as well as the measured \cite{Jia:2012ve} event-by-event distribution of $v_2$, $v_3$, and $v_4$, again for a range of collision centralities \cite{Gale:2012rq}. This is shown in Fig.~\ref{F7} for Pb+Pb collisions at the LHC but also holds for Au+Au collisions at RHIC \cite{Gale:2012rq}. The only difference is that for LHC energies the effective specific shear viscosity of the QGP fluid must be chosen somewhat larger ($(\eta/s)_\mathrm{LHC}{\,\simeq\,}0.2$) than at top RHIC energy ($(\eta/s)_\mathrm{RHIC}{\,\simeq\,}0.12$). Both data sets are consistent with a temperature dependent specific shear viscosity $(\eta/s)(T)$ proposed in \cite{Niemi:2011ix} (the thin lines in the lower left panel in Fig.~\ref{F7}) that has a minimum value around $1/(4\pi){\,=\,}0.08$ at the pseudocritical temperature $T_\mathrm{cr}$ where the QGP hadronizes and then rises to about 5 times that minimal value at $2T_\mathrm{cr}$. While this temperature-dependence of $\eta/s$ is not yet tightly constrained by the available analyses of RHIC and LHC data, the need for a somewhat larger effective shear viscosity at the higher temperatures probed at the LHC, already noted in earlier work \cite{Luzum:2010ag,Niemi:2011ix}, appears to solidify, and the early evidence from RHIC data for a very low specific shear viscosity near $T_\mathrm{cr}$ \cite{Lacey:2006pn}, perhaps as low as the KSS bound \cite{Policastro:2001yc} of $1/(4\pi)$, is strongly supported by the work in \cite{Gale:2012rq}.

\vspace*{-3mm}
\section{Summary: status of and prospects for the Little Bang Standard Model}
\label{sec4}

The Little Bang Standard Model is still under construction, but its key features are showing through the scaffolding: {\bf 1.}\ Every heavy-ion collision system, centrality class, and collision energy generates a different class of Little Bangs, each with its own characteristic initial density fluctuation and final anisotropic flow fluctuation spectrum. {\bf 2.}\ The initial fluctuation spectrum can be computed from the Color Glass Condensate (CGC) theory using e.g. the IP-Glasma model; gluon field fluctuations inside the nucleons within the colliding nuclei play an essential role in this spectrum. {\bf 3.}\ After a very short pre-equilibrium evolution stage, not lasting much longer than 1\,fm/$c$, which is best described using classical Yang-Mills field dynamics corrected for quantum fluctuations, the Little Bangs are in the QGP phase which undergoes viscous hydrodynamic evolution until it has cooled down to the pseudo-critical temperature for hadronization, $T_\mathrm{cr}{\,\simeq\,}160$\,MeV. The specific QGP shear viscosity is small, around $1/(4\pi)$ near $T_\mathrm{cr}$, rising modestly at higher temperatures; its effective value for 200\,$A$\,GeV Au+Au collisions at RHIC is around $0.12{\,=\,}1.5/(4\pi)$, for 2.76\,$A$\,TeV Pb+Pb collisions it is around $0.2{\,=\,}2.5/(4\pi)$. This small shear viscosity allows flow anisotropies to build in response to the initial fluctuations and anisotropies in the pressure gradients, to values that are large enough to be experimentally observed in the final state but still visibly attenuated by shear viscous effects, following an $\eta/s$-dependent pattern where higher harmonic flow coefficients are suppressed more strongly than lower harmonics. The spectrum of flow anisotropies, their magnitudes, directions, particle species and $\pT$-dependences, and event-by-event fluctuations provide a rich menu of experimental observables from which the QGP shear viscosity and initial fluctuation spectra can be reconstructed. {\bf 4.}\ After hadronization, the Little Bang continues to evolve as a dilute, highly dissipative hadron resonance gas that is highly inefficient in generating any additional momentum anisotropies but also doesn't erase those established earlier during the QGP phase. This makes it possible to extract from hadronic final state observables quantitative information about the earlier QGP and CGC phases. 

In this relatively simple form, the Little Bang Standard Model applies to Little Bangs created at top RHIC and LHC energies. At lower collision energies, our understanding of the initial conditions becomes less reliable and the lifetime of the hydrodynamic QGP stage shrinks, making a quantitative theory of the Little Bang more challenging, but not less interesting.

\ack
This work was supported by the U.S.\ Department of Energy under Grants No.~\rm{DE-SC0004286} and (within the framework of the JET Collaboration) \rm{DE-SC0004104}. I thank my (former) students Huichao Song, Chun Shen, and Zhi Qiu for their important contributions to the construction of the LBSM.


\section*{References}
\begin{thebibliography}{99}

\bibitem{Schenke:2012wb} 
  Schenke B, Tribedy P, and Venugopalan R 2012
  {\it Phys.\ Rev.\ Lett.}  {\bf 108} 252301; 
  and 
  {\it Phys.\ Rev.\ C} {\bf 86} 034908

\bibitem{NSAC}
  Tribble R (chair), Burrows A {\it et al.} 2013 {\sl Implementing the 2007 Long Range Plan}, 
  Report to the Nuclear Science Advisory Committee, January 31, 2013. Available at 
  \url{http://science.energy.gov/np/nsac/reports/ .}
  
\bibitem{Planck:2013kta} 
  Ade PAR {\it et al.} (Planck Collab.) 2013
  {\it Preprint} 1303.5075 [astro-ph.CO]

\bibitem{Heinz:2013th} 
  Heinz U and Snellings R 2013
  {\it Annu. Rev. Nucl. Part. Sci.} {\bf 63} in press ({\it Preprint} 1301.2826 [nucl-th]).

\bibitem{Heinz:1998nk} 
  Heinz U 1999
  in {\sl Strong and Electroweak Matter '98}, J. Ambjorn et al. (eds.),
  World Scientific, Singapore, p. 81-100 ({\it Preprint} arXiv:hep-ph/9902424);
  and 
  {\it Nucl.\ Phys.}\ {\bf A661} 140 

\bibitem{Shen:2011eg} 
  Shen C, Heinz U, Huovinen P, and Song H 2011
  {\it Phys.\ Rev.\ C} {\bf 84} 044903
  
\bibitem{CC}
  Harris JW, Kharzeev D, and Ullrich T 2012 {\it CERN Courier} {\bf 52} (no. 9) 17 

\bibitem{Kolb:2003dz} 
  Kolb PF and Heinz U 2004
  in {\sl Quark-Gluon Plasma 3}, Hwa RC {\it et al.} (eds.), World Scientific,  
  Singapore, p. 634-714 ({\it Preprint} arXiv:nucl-th/0305084).

\bibitem{Heinz:2001xi} 
  Heinz U and Kolb PF 2002
  {\it Nucl.\ Phys.}\ {\bf A702} 269

\bibitem{Kovner:1995ja} 
  Kovner A, McLerran LD, and Weigert H 1995
  {\it Phys.\ Rev.\ D} {\bf 52} 6231;
  Kovchegov YV and Rischke DH 1997
  {\it Phys.\ Rev.\  C} {\bf 56} 1084;
  Krasnitz A and Venugopalan R 1999
  {\it Nucl.\ Phys.}\ {\bf B557} 237;
  Krasnitz A and Venugopalan R 2000
  {\it Phys.\ Rev.\ Lett.}\ {\bf 84} 4309;
  Krasnitz A and Venugopalan R 2001
  {\it Phys.\ Rev.\ Lett.} {\bf 86} 1717;
  Lappi T 2003
  {\it Phys.\ Rev.\ C} {\bf 67} 054903;
  Lappi T and McLerran LD 2006
  {\it Nucl.\ Phys.}\ {\bf A772} 200

\bibitem{Gale:2012rq} 
  Gale C, Jeon S, Schenke B, Tribedy P, and Venugopalan R 2013
  {\it Phys. Rev. Lett.} {\bf 110} 012302;
  and
  {\it Preprint} 1210.5144 [hep-ph]

\bibitem{Heinz:2013bua} 
  Heinz U, Qiu Z, and Shen C 2013
  {\it Phys. Rev. C} {\bf 87} 034913
  
\bibitem{Jia:2012ma} 
  Jia J and Mohapatra S 2012
  {\it Preprint} 1203.5095 [nucl-th];
  Jia J and Teaney D 2012
  {\it Preprint} 1205.3585 [nucl-ex]
  
\bibitem{Qiu:2012uy} 
  Qiu Z and Heinz U 2012
  {\it Phys.\ Lett.\ B} {\bf 717} 261

\bibitem{ALICE:2011ab} 
  Aamodt  K {\it et al.} (ALICE Collab.) (2011)
  {\it Phys.\ Rev.\ Lett.} {\bf 107} 032301;
  Bilandzic A {\it et al.} (ALICE Collab.) 2012
  {\it Preprint} 1210.6222 [nucl-ex]

\bibitem{Jia:2012sa} 
  Jia J {\it et al.} (ATLAS Collab.) 2012
  {\it Preprint} 1208.1427 [nucl-ex]
  
\bibitem{Qiu:2011iv} 
  Qiu Z and Heinz U 2011
  {\it Phys.\ Rev.\ C} {\bf 84} 024911

\bibitem{Teaney:2012ke} 
  Teaney D and Yan L 2012
  {\it Phys.\ Rev.\ C} {\bf 86}, 044908;
  and
  {\it Preprint} 1210.5026 [nucl-th]

\bibitem{Ollitrault:2009ie}
  Ollitrault J-Y, Poskanzer AM, and Voloshin SA 2009
  {\it Phys.\ Rev.\  C} {\bf 80} 014904

\bibitem{Adams:2004bi}
  Adams  J {\it et al.} (STAR Collab.) 2005
  {\it Phys.\ Rev.\  C} {\bf 72} 014904

\bibitem{Song:2010aq} 
  Song H, Bass SA, and Heinz U 2011
  {\it Phys.\ Rev.\ C} {\bf 83} 024912;
  Song H, Bass SA, Heinz U, Hirano T, and Shen C 2011
  {\it Phys.\ Rev.\ Lett.}\ {\bf 106}, 192301

\bibitem{Schenke:2011bn} 
  Schenke B, Jeon S, and Gale C 2012
  {\it Phys.\ Rev.\ C} {\bf 85} 024901

\bibitem{ATLAS:2012at} 
  Aad G {\it et al.} (ATLAS Collab.) 2012
  {\it Phys.\ Rev.\ C} {\bf 86} 014907

\bibitem{Jia:2012ve} 
  Jia J {\it et al.} (ATLAS Collab.) 2012
  {\it Preprint} 1209.4232 [nucl-ex];
  ATLAS Collab 2012 {\it ATLAS note ATLAS-CONF-2012-114.}
  \url{http://cds.cern.ch/record/1472935/files/ATLAS-CONF-2012-114.pdf}

\bibitem{Song:2011hk} 
  Song H, Bass SA, Heinz U, Hirano T, Shen C 2011
  {\it Phys.\ Rev.\ C} {\bf 83} 054910

\bibitem{Song:2009rh} 
  Song H and Heinz U 2010
  {\it Phys.\ Rev.\ C} {\bf 81} 024905
  
\bibitem{Heinz:2011kt} 
  Heinz U, Shen C, and Song H 2012
  {\it AIP Conf.\ Proc.}\ {\bf 1441} 766

\bibitem{Abelev:2012wca} 
  Abelev B {\it et al.}  (ALICE Collab.) 2012
  {\it Phys.\ Rev.\ Lett.}\  {\bf 109} 252301;
  and 2013 {\it Preprint} 1303.0737 [hep-ex]

\bibitem{Muller:2012zq} 
  Muller B, Schukraft J, and Wyslouch B
  {\it Annu.\ Rev.\ Nucl.\ Part.\ Sci.}\  {\bf 62} 361

\bibitem{Qiu:2011hf} 
  Qiu Z, Shen C, and Heinz U (2012)
  {\it Phys.\ Lett.\ B} {\bf 707} 151

\bibitem{Schenke:2012hg} 
  Schenke B, Tribedy P, and Venugopalan R 2012
  {\it Phys.\ Rev.\ C} {\bf 86} 034908 

\bibitem{Bartels:2002cj} 
  Bartels J, Golec-Biernat KJ, and Kowalski H (2002)
  {\it Phys.\ Rev.\ D} {\bf 66} 014001;
  Kowalski H and Teaney D 2003
  {\it Phys.\ Rev.\ D} {\bf 68} 114005

\bibitem{Tribedy:2010ab} 
  Tribedy P and Venugopalan R 2011
  {\it Nucl.\ Phys.}\ {\bf A850} 136

\bibitem{Niemi:2011ix}
  Niemi H, Denicol GS, Huovinen P, Molnar E, and Rischke DH 2011
  {\it Phys.\ Rev.\ Lett.}\ {\bf 106} 212302;
  and 2012 {\it Phys.\ Rev.\ C} {\bf 86} 014909
  
\bibitem{Luzum:2010ag} 
  Luzum M 2011
  {\it Phys.\ Rev.\ C} {\bf 83} 044911;
  Song H, Bass SA, and Heinz U 2011
  {\it Phys.\ Rev.\ C} {\bf 83} 054912

\bibitem{Lacey:2006pn} 
  Lacey RA and Taranenko A 2006
  {\it Proceedings of Science} {\bf CFRNC2006} 021
  
\bibitem{Policastro:2001yc} 
  Policastro G, Son DT, and Starinets AO 2001
  {\it Phys.\ Rev.\ Lett.}\  {\bf 87} 081601;
  Kovtun P, Son DT, and Starinets AO 2005
  {\it Phys.\ Rev.\ Lett.}\ {\bf 94} 111601
  
  
\end{thebibliography}
\end{document}